\documentclass[twocolumn,twocolappendix]{aastex631}
\usepackage{upgreek}
\usepackage{amsmath,amssymb,amsfonts,bm}
\usepackage{epstopdf}
\usepackage{epsfig}
\usepackage{natbib,caption2}
\usepackage{graphicx}   
\usepackage{float}
\usepackage{longtable}
\usepackage{graphics}
\usepackage{hyperref}
\usepackage{color}
\usepackage{calc}
\usepackage{threeparttable}
\usepackage{ulem}
\usepackage{booktabs}
\usepackage{url}

\newcommand \beq{\begin{equation}}
\newcommand \eeq{\end{equation}}
\newcommand \bey{\begin{eqnarray}}
\newcommand \eey{\end{eqnarray}}

\newcommand \kms{\,{\rm km \, s}^{-1}}

\newcommand{\gsim}{\lower.5ex\hbox{$\; \buildrel > \over \sim \;$}}
\newcommand{\lsim}{\lower.5ex\hbox{$\; \buildrel < \over \sim \;$}}
\newcommand{\ha}{\hbox{H$\alpha$}}

\newcommand{\oii}{\hbox{[O\,{\sc ii}]}}

\newcommand  \siiv  {\ifmmode {\rm Si}\, {\sc iv}\ \else Si\,{\sc iv}\fi}
\newcommand  \SIIV  {\ifmmode {\rm Si}\,{\sc iv}\,\lambda1399 \else Si\,{\sc iv}\,$\lambda1399$\fi}
\newcommand  \civ  {\ifmmode {\rm C}\, {\textsc iv}\ \else C\,{\sc iv}\fi}
\newcommand  \CIV  {\ifmmode {\rm C}\,{\sc iv}\,\lambda1549 \else C\,{\sc iv}\,$\lambda1549$\fi}
\newcommand{\civab}{C{\sc~iv}\,$\lambda\lambda$1548,1551}
\newcommand{\siivab}{Si{\sc~iv}\,$\lambda\lambda$1393,1402}
\newcommand  \aliii  {\ifmmode {\rm Al}{\textsc{iii}} \else Al\,{\sc iii}\fi}
\newcommand  \ALIII  {\ifmmode {\rm Al}\,{\sc iii}\,\lambda1854 \else Al\,{\sc iii}\,$\lambda1854$\fi}
\newcommand  \feii {\ifmmode {\rm Fe}\,{\textsc{ii}}\, \else Fe\,{\sc ii}\fi}

\def\mgii{Mg\,{\sc~ii}}
\newcommand  \MGII  {\ifmmode {\rm Mg}\,{\sc ii}\,\lambda2798 \else Mg\,{\sc ii}\,$\lambda2798$\fi}

\newcommand{\mgiiab}{Mg{\sc~ii}\,$\lambda\lambda$2796,2803}

\def\kms{$\rm km\,s^{-1}$}

\acceptjournal{ApJ}

\shortauthors{Chen et al.}

\begin{document}

\title{The infrared traced star formation rates of associated absorption lines quasars}

\correspondingauthor{Rui-Jing Lu}
\email{luruijing@gxu.edu.cn}

\author{Zhe-Geng Chen}
\affiliation{Guangxi Key Laboratory for Relativistic Astrophysics, School of Physical Science and Technology, Guangxi University, Nanning 530004,People's Republic of China}
\affiliation{School of Physics and Electronic Information, Guangxi Minzu University, Nanning 530006, People's Republic of China}

\author[0000-0003-2467-3608]{Rui-Jing Lu}
\affiliation{Guangxi Key Laboratory for Relativistic Astrophysics, School of Physical Science and Technology, Guangxi University, Nanning 530004,People's Republic of China}

\author{Zhi-fu Chen}
\affiliation{School of Physics and Electronic Information, Guangxi Minzu University, Nanning 530006, People's Republic of China}

\author{Wen-Qiang Liang}
\affiliation{Guangxi Key Laboratory for Relativistic Astrophysics, School of Physical Science and Technology, Guangxi University, Nanning 530004,People's Republic of China}

\author{Xing-long Peng}
\affiliation{School of Physics and Electronic Information, Guangxi Minzu University, Nanning 530006, People's Republic of China}

\author{Jing Li}
\affiliation{Guangxi Key Laboratory for Relativistic Astrophysics, School of Physical Science and Technology, Guangxi University, Nanning 530004,People's Republic of China}

\author{Wei-rong Huang}
\affiliation{School of Physics and Electronic Information, Guangxi Minzu University, Nanning 530006, People's Republic of China}




\begin{abstract}
Some optically selected quasars exhibit \mgii\ assoicated absorption lines (AALs), and its origin remains unclear. In this paper, we compile a sample of 1769 quasars, with or without \mgii\ AALs. Of which 1689 are Far-Infrared (FIR) detected quasars and the rest are not detected in FIR. For the FIR undetected quasars, we obtain stacks for both with and without \mgii\ AAL quasars. Then we estimate the star formation rates (SFRs) within quasar host galaxies based on their FIR luminosities derived from their FIR greybody components, and find that, although quasars with \mgii\ AALs have significantly redder median composite spectra than those without \mgii\ AALs, the SFR distributions of the two types of quasars are statistically indistinguishable. These results do not require an evolutionary link between the quasars with and without \mgii\ AALs, and would be reconciled if an orientation effect cannot be ignored among the quasars hosting different types of absorption lines.

\end{abstract}


\section{Introduction}

The star formation rate (SFR) is a crucial astrophysical tracer for understanding the formation and evolution of galaxies \citep[][]{2019ApJ...882...89Z}. In the evolutionary process of galaxies, active galactic nucleus (AGNs) could be triggered by gas-rich galaxy mergers, leading to black hole accretion and bursts of star formation in host galaxies \citep[][]{1988ApJ...325...74S, 2008ApJS..175..356H, 2016MNRAS.458..816H}. A substantial amount of gas and dust is required to funnel inward to fuel black hole accretion and star-forming activities, which also contributes to the obscuration of young AGNs. As the central black hole grows, the jet and wind, which originate in the central regions of AGNs, subsequently generate feedback to surrounding environments, potentially enhancing or quenching the star formation rate (SFR) within host galaxies \citep[e.g.,][]{2012ARA&A..50..455F,2012MNRAS.425..438G,2016MNRAS.455.4166B,2022NatAs...6..339C}.

The gas clouds that are far beyond the gravitational bound of quasar systems often produce absorption lines with redshifts much less than those of quasar systems ($z_{abs}\ll z_{em}$) in quasar spectra, which are generally called intervening absorption lines and often host line widths less than a few hundred \kms\ (narrow absorption lines, NALs). Whilst, the gas medium within the gravitational bound of quasar systems often produces absorption lines with $z_{abs} \approx z_{em}$, which are often called as associated absorption lines (AALs) and have a wide range of line widths. Thus the AALs can be further divided into three subclasses based on their line widths: (1) broad absorption lines (BALs) with line widths generally larger than 2000 \kms; (2) NALs; and (3) Mini-BALs with line widths between the BALs and NALs. Meanwhile, according to the ionization levels of transition lines, the AALs can be categorized into high-ionization AALs (HiAALs, e.g., \siivab; \civab) and low-ionization AALs (LoAALs, e.g., \mgiiab; \ALIII).

About 30\% of optically selected quasars exhibit AALs \citep[e.g.,][]{2009ApJ...692..758G,2011MNRAS.410..860A,2016MNRAS.462.2980C,2018ApJS..235...11C,2024ApJ...963....3P}. Therefore, AALs serve as a useful tool to probe the physical environments within quasar systems.

Some studies have shown that quasars with LoAALs (including LoBALs, LoMini-BALs, and LoNALs) have significantly higher SFR than those without AALs (Non-AALs) \citep[e.g.,][]{2022NatAs...6..339C,2024ApJ...963....3P}.

Whereas, different results have also been reported that quasars with high-ionization broad absorption lines (HiBALs) exhibit a SFR consistent with those without BALs (Non-BALs) \citep[e.g.,][]{2004ASPC..311..223W, 2012MNRAS.427.1209C, 2016MNRAS.462.4067P}.

There are main two schemes for explaining whether the quasars host AALs: the orientation-dependent and evolution schemes. In the evolution scheme, the quasars hosting AALs are often considered to live in the gas-rich environment of the early stages of galaxy evolution \citep[e.g.,][]{1992ApJ...399L..15B,2016ApJ...824..106W,2016ApJ...820..121B,2022NatAs...6..339C}.
Consequently, the AAL quasar often host intense star formation activity and thus has high dust masses which will yield higher FIR emission. When the quasar evolves from the AALs to the Non-AALs, the SFRs within host galaxies naturally reduce if the powerful outflows of quasars sweep out the interstellar gas. In the orientation-dependent hypothesis \citep[e.g.,][]{1993ARA&A..31..473A,1995PASP..107..803U,2012ASPC..460...47H}, AALs are likely to appear in spectra if the quasar's line of sight is close to the accretion disk, otherwise they are unlikely to appear. Consequently, the SFR within the host galaxy does not depend on whether the quasar has AALs.

The UV-optical tracers of the SFR, such as \oii, \ha\ emission lines \citep[e.g.,][]{2012ApJ...748..131S,2014ARA&A..52..415M,2020ApJ...893...25C,2024ApJ...963....3P}, are susceptible to the dust attenuation and the contamination from quasar central regions \citep[e.g.,][]{1998ARA&A..36..189K, 2000ApJ...533..682C, 2020MNRAS.493.3966K}. To better assess the SFR within the host galaxy, many previous efforts have mainly focused on corrections for both the dust attenuation and the contamination from quasar emissions \citep[e.g.,][]{2019ApJ...882...89Z}. However, these corrections may also invoke other biases. The dust attenuation at FIR wavelength is generally considered as negligible. Consequently, the FIR emission is an important tracer of the SFR within host galaxy. Using the FIR tracer of the SFR, \citet{2012MNRAS.427.1209C} claimed that the HiBAL quasars exhibit the similar SFRs  with the Non-BAL ones, which may be partly due to the fact that the HiBAL quasars are at an evolutionary stage close to the Non-BAL quasars. The evolutionary stage of the LoAAL quasars, including LoBALs, LoMini-BALs, and LoNALs, is likely far away from that of the Non-AAL quasars \citep[e.g.,][]{2022NatAs...6..339C,2024ApJ...963....3P}. Therefore, we compile a sample of quasars with FIR observations to compare the SFR within host galaxies between the LoAAL and Non-LoAAL quasars.

Throughout this work, we assume a flat $\Lambda$CDM cosmology with $\Omega_m$ = 0.3,  $\Omega_\Lambda$ = 0.7, and $h_0$ = 0.7.

\section{Data and analysis} \label{sec:2}
\subsection{Quasar sample}

The Sloan Digital Sky Survey \citep[SDSS;][]{2000AJ....120.1579Y} collected quasar spectra at a resolution $R\approx2000$ in wavelength ranges of $\lambda \approx 3800-9200$ {\AA} \citep[SDSS-I/II][]{2009ApJS..182..543A}, or  $\lambda \approx 3600-10500$ {\AA} \citep[SDSS-III/IV][]{2013AJ....146...32S,2013AJ....145...10D}. The Sixteenth Data Release(DR16Q) is the final dataset for the SDSS-IV quasar catalog \citep[][]{2020ApJS..250....8L}, which contains 750,414 quasars. In this work, we aim to study quasars with and without \mgiiab\ AALs. Therefore, to ensure reliable detection of  \mgiiab\ absorption lines in the SDSS spectra, if any, we select the quasars with $0.3<z_{\rm em}<2.0$ from the SDSS DR16Q.

To obtain the integrated FIR luminosities from the spectral energy distributions (SED) of quasars, we collect their corresponding photometric data at mid-infrared(MIR) 3.4 $\rm \upmu m$ (W1), 4.6 $\rm \upmu m$ (W2), 12 $\rm \upmu m$ (W3), and 22 $\rm \upmu m$ (W4) from the data catalog of the Wide-field Infrared Survey Explorer \citep[WISE;][]{2018ApJS..234...23A}, and at FIR 250 $\rm\upmu m$, 350 $\rm\upmu m$, and 500 $\rm\upmu m$ from the \textit{Herschel} catalog \citep[][]{2014ApJS..210...22V,2020ipac.data..I45N,2020ipac.data..I46N,2020ipac.data..I44N}. Due to the fact that most of the SDSS quasars are not detected in the W4 band \citep[e.g.,][]{2019ApJ...871..136B,2020ApJS..250....8L}, and given that during the WISE mission, the W4 band became unusable due to the depletion of its solid hydrogen cryogen, the W1, W2, and W3 bands continued with another half-sky scan. Thus, the data from the W4 band are excluded from our analysis.
We initially arrive at a total of 1779 quasars, all having the 3.4 $\rm \upmu m$, 4.6 $\rm \upmu m$, 12 $\rm \upmu m$, 250 $\rm\upmu m$, 350 $\rm\upmu m$, and 500 $\rm\upmu m$ photometric data. 90 of 1779 quasars have flux densities under the nominal 3$\sigma$ depths of 17.4, 18.9, and 20.4 $\rm mJy$ at 250, 350, and 500 $\rm \upmu m$, respectively \citep[][]{2010A&A...518L...5N,2016ApJ...819..123N}, which are classified into FIR undetected quasars. Among the FIR undetected quasars, 10 are excluded because they are located too close to the edge of their respective fields for further stacking analysis. We finally arrive at a total of 1769 quasars.
Of these, 1689 are FIR detected quasars, and 80 are FIR undetected ones.

\begin{figure}[ht!]
	\includegraphics[width=0.48\textwidth]{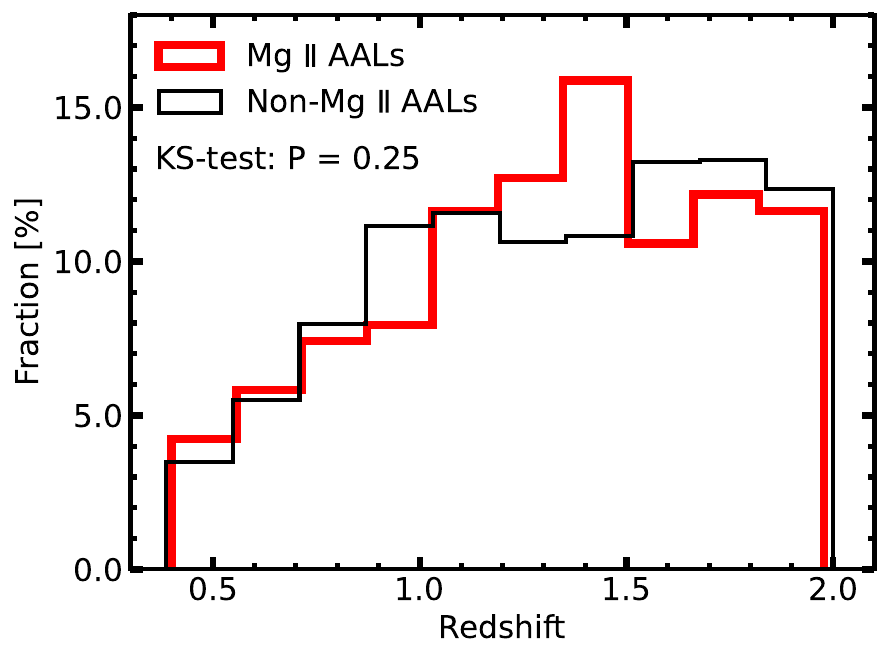}\\
	\caption{Redshift distributions for the quasars with and without \mgii\ AALs. The KS test yields a probability $p= 0.25$.}
\label{fig:z}
\end{figure}

Based on the SDSS DR16Q catalog, previous works \citep{2018ApJS..235...11C,2022NatAs...6..339C,2024ApJ...963....3P} have systematically searched for the \mgii\ AALs within the velocity offset $|\upsilon_{\rm off}|$\footnote{$\upsilon_{\rm off} = \frac{(1+z_{\rm abs})^2 - (1+z_{\rm em})^2}{(1+z_{\rm abs})^2 + (1+z_{\rm em})^2}$}  of 3000 \kms\ around \mgii\ emission lines. Following the same method, in order to work on a well-defined, high-quality sample of quasars with and without \mgii\ AALs, in this paper, we construct a sample of quasars with \mgii\ AALs within $|\upsilon_{\rm off}|<1500$ \kms\ (\mgii\ AAL quasars hereafter), and sample of quasars without \mgii\ AALs $|\upsilon_{\rm off}|<6000$ \kms\ (Non-\mgii\ AAL quasars hereafter).
Of the 1769 quasars, 189 have \mgii\ AALs, and 1580  do not. Of the 189 AAL quasars, 182 have the FIR detections and 7 do not. While, Of the 1580 Non-AAL quasars, 1507 have the FIR detections, and 73 do not. As shown in Figure \textcolor{blue}{\ref{fig:z}}, the two types of quasars share almost the same redshift distribution with a median redshift of 1.35.

\subsection{SED fitting and FIR SFR}\label{SED fitting and FIR SFR}
The IR emission of galaxies comes mainly from the combined contribution of older stellar populations, AGNs, and star formation (SF) \citep[e.g.,][]{2014ARA&A..52..415M, 2008ApJ...672..214G, 2015A&A...576A..10C}, in which, the AGN contribution is primarily originated in the dusty environment (the dust in the torus and clouds), which drops sharply at longer wavelengths, and negligible at $\lambda\gtrsim 30~\upmu$m \citep[e.g.,][]{2012MNRAS.420..526M,2016MNRAS.459..257S,2017MNRAS.471...59L}. Meanwhile, the older stellar population contribution  is dominated by the radiations at $\lambda \lesssim 2~\upmu$m. Thus, this contribution to the FIR luminosity is also negligible \citep[e.g.,][]{2016MNRAS.462.4067P, 2020MNRAS.498.1469W}. The SF contribution to IR emission mainly originates from the re-radiation of dust, which absorbed the UV emission primarily from star formation. This contribution is characterized by a greybody model \citep{2012MNRAS.425.3094C}:

\begin{equation} S(\lambda,T)\propto(1-e^{-\left[\frac{\lambda_0}{\lambda}\right]^\beta})\frac{1}{\lambda^5}\frac{1}{e^{hc/{\lambda}kT}-1},
\end{equation}
where $\lambda_0$ represents the wavelength at which optical depth reaches unity \citep{2006ApJ...636.1114D}, $T$ is the dust temperature constrained within 15--60 K \citep[e.g.,][]{2014ARA&A..52..415M,2020MNRAS.498.1469W},
spanning the full range of observed dust temperatures for quasar hosts of our sample, and $\beta$ is the emissivity index with a fixed value of $\beta$=1.6 \citep{2001MNRAS.324L..17P}.

\begin{figure}[ht!]
\includegraphics[width=0.48\textwidth]{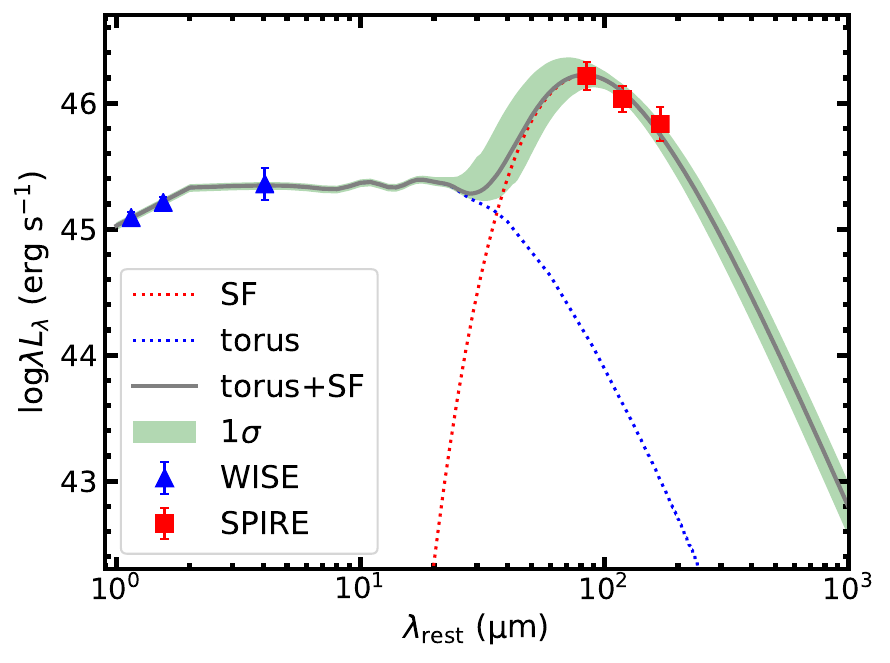}\\
\caption{The best SED fit to the quasar $J094729.14+381033.2$. Blue squares represent the WISE photometries at 3.4 $\upmu$m, 4.6 $\upmu$m, and 12 $\upmu$m. Red squares represent the \textit{Herschel} photometries at 250 $\upmu$m, 350 $\upmu$m, and 500 $\upmu$m. The blue-dotted line represents the contribution from the AGN \citep[][]{2017MNRAS.471...59L}, the red-dotted line represents the contribution from the greybody emission \citep{2012MNRAS.425.3094C}, and grey-solid line is the sum of the blue-dotted and red-dotted lines. Green-shaded region marks 68\% credible interval of the best model.}		
\label{fig:sed}
\end{figure}

As analyzed above, the IR emission of quasars could simultaneously fit a total model with a combination of the two components, i.e.
\begin{equation}\label{eq:total_model}
M_{\text{tot}}=X_{\text{torus}} \times M_{\text{torus}}({\lambda}) + X_{\text{SF}} \times S(\lambda),
\end{equation}
where the $M_{\rm torus}$ represents the AGN contribution, which was derived from 115 type-I AGNs \citep{2017MNRAS.471...59L}, and the two components of $M_{\text{torus}}$ and $S(\lambda)$ are  scaled by the factors of $ X_{\text{torus}}$ and $ X_{\text{SF}}$, respectively. Throughout the fitting, we take the two vertical scalings and the dust temperature of the greybody template as free parameters, with 6 data points for each of quasars in our sample. In order to obtain full posterior distributions on the best-fitting model parameters and to marginalize over any nuisance parameters, we utilize Bayesian approach and Markov chain Monte Carlo (MCMC) sampling based on the Python package {\tt emcee} \citep{2013PASP..125..306F}, and finally get the best dust temperature $T$, $X_{\text{torus}}$ and $X_{\text{SF}}$, respectively. As an example, Figure \textcolor{blue}{\ref{fig:sed}} shows the best fit model to the observation of quasar $\rm J094729.14+381033.2$. With the best fitting parameters, we obtain the FIR luminosity ($L_{\rm FIR}$) of quasar by integrating the greybody curve (see the red-dotted line of Figure \textcolor{blue}{\ref{fig:sed}}) from 8 to 1000 $\mu$m . Thus the SFR within its host galaxy can be estimated with the original empirical relation derived by \cite{1998ARA&A..36..189K}:

\begin{equation}
	\rm{SFR(M_\odot~yr^{-1})} = 4.5 \times 10^{-44} \mathit{L}  \rm{_{FIR}~(erg~s^{-1})},
\label{con:Kenni}
\end{equation}
that assumes a continuous starbursts of age 10 --- 100 Myr. For the FIR detected quasars included in our final sample, the resulting SFR are listed in Table \textcolor{blue}{\ref{tab:properties}}.

\begin{table*}[htbp]
	\centering
	\caption{The properties of the sample}
	\tabcolsep 0.6mm
    \footnotesize\tiny
	\label{tab:properties}
	\begin{tabular}{ccccccccccccccc}
		\hline\hline\noalign{\smallskip}
		SDSS name & PLATE &  MJD & FIBERID & z & $S_{3.4}$ & $S_{4.6}$ & $S_{12.0}$ & $S_{250}$ & $S_{350}$ & $S_{500}$ & $\log M_{\rm{BH}}$  & SFR & AAL & FIR\\
		&&&&&mJy&mJy&mJy&mJy&mJy&mJy& $ \rm M_\odot$  & $ \rm M_\odot~\rm{yr} ^{-1}$ &&detected?\\
		\hline\noalign{\smallskip}
		000022.44-031041.5  &7895& 57659 & 918	&1.2011 &0.031$\pm$0.005 &0.065$\pm$0.011 &0.049$\pm$0.129 &64.0$\pm$19.7 &79.2$\pm$18.0 &73.0$\pm$17.9 &	8.45 &	$330_{-49}^{+68}$   &	0&   yes\\
		000055.41-012006.0  &7850& 56956 & 734	&1.7581 &0.034$\pm$0.005 &0.050$\pm$0.010 &0.151$\pm$0.111 &32.0$\pm$6.2 &38.2$\pm$6.8 &24.2$\pm$6.6 &	8.67 &	$397_{-51}^{+80}$   &	0&   yes\\
		000059.09-013722.6  &7850& 56956 & 280	&1.6835 &0.066$\pm$0.005 &0.083$\pm$0.010 &0.020$\pm$0.125 &50.2$\pm$8.3 &43.9$\pm$10.2 &33.2$\pm$7.9 &	8.84 &	$550_{-84}^{+130}$  &	0&   yes\\
		000100.36-021237.3  &7850& 56956 & 266	&1.2979 &0.033$\pm$0.005 &0.089$\pm$0.011 &0.036$\pm$0.134 &82.7$\pm$15.8 &82.7$\pm$20.8 &67.1$\pm$20.7 &	8.53 &	$473_{-73}^{+96}$   &	0&   yes\\
		000122.41-000130.6  &7848& 56959 &  18 	&1.4830 &0.150$\pm$0.005 &0.238$\pm$0.010 &0.178$\pm$0.110 &91.0$\pm$13.9 &92.7$\pm$16.8 &59.0$\pm$16.9 &	8.01 &	$733_{-89}^{+165}$  &   1&   yes \\
		\noalign{\smallskip}
		\hline\hline\noalign{\smallskip}
	\end{tabular}
	\begin{flushleft}
		Note --- Numbers of ``0''  and ``1'' in column `AAL' represent the spectra without or with \mgii\  AALs.\\
(This table is available in its entirety in machine-readable form.)
	\end{flushleft}
\end{table*}

\subsection{Stacks of the FIR undetected quasars}\label{stacking}
Stacking FIR images based on optically measured positions of FIR undetected quasars could reduce confusion, thereby enhances the accuracy of the mean flux density measurements \citep[e.g.,][]{2016MNRAS.458.1478C, 2017MNRAS.472.2221S, 2019ApJ...871..136B}. To yield the mean flux densities at 250 $\rm\upmu m$, 350 $\rm\upmu m$, and 500 $\rm\upmu m$ for the FIR undetected quasars, we here stack the FIR images observed by \textit{Herschel} based on the optically measured positions \citep[e.g.,][]{2016MNRAS.462.3146V}. We briefly describe the main processes as follows. We cut the image around each FIR undetected quasar with a set size of $5' \times 5'$, and stack the cutting images using the inverse of the noise as the weight. A combination of a 2D Gaussian plus a constant background is invoked to fit the stacked image. The resulting peak value is taken as the mean flux density. The corresponding uncertainty is yielded with the bootstrap technique. Using the same method shown in Section \textcolor{blue}{\ref{SED fitting and FIR SFR}}, we also estimate their SFRs and yield $73^{+41}_{-25}\rm{~M_\odot~yr^{-1}}$ and $65^{+16}_{-13}\rm{~M_\odot~yr^{-1}}$ for the quasars with and without \mgii\ AALs, respectively, which indicates that there is no statistically significant difference in SFRs of the two types of FIR undetected quasars.

\section{Discussions and conclusions} \label{sec:DISCUS}
We have compiled a sample of 1769 quasars with $0.3<z_{\rm em}<2.0$ from the SDSS DR16Q, among which 189 quasars are detected \mgii\ AALs within $|\upsilon_{\rm off}|<1500$ \kms, and 1580 quasars are without detected \mgii\ AALs within the spectral data within $|\upsilon_{\rm r}|<6000$ \kms. All of these 1769 quasars have been observed by the WISE at 3.4 $\rm \upmu m$, 4.6 $\rm \upmu m$, 12 $\rm \upmu m$, and by the \textit{Herschel} at 250 $\rm\upmu m$, 350 $\rm\upmu m$, and 500 $\rm\upmu m$.
Among the 189 quasars with \mgii\ AALs, \textit{Herschel} FIR flux densities of 7 quasars are below the 3$\sigma$ threshold, which are classified as FIR undetected quasars. Among the 1580 quasars without \mgii\ AALs, 73 quasars are below the 3$\sigma$ threshold, which are also classified as FIR undetected quasars.

\begin{figure}[ht!]
	\includegraphics[width=0.49\textwidth]{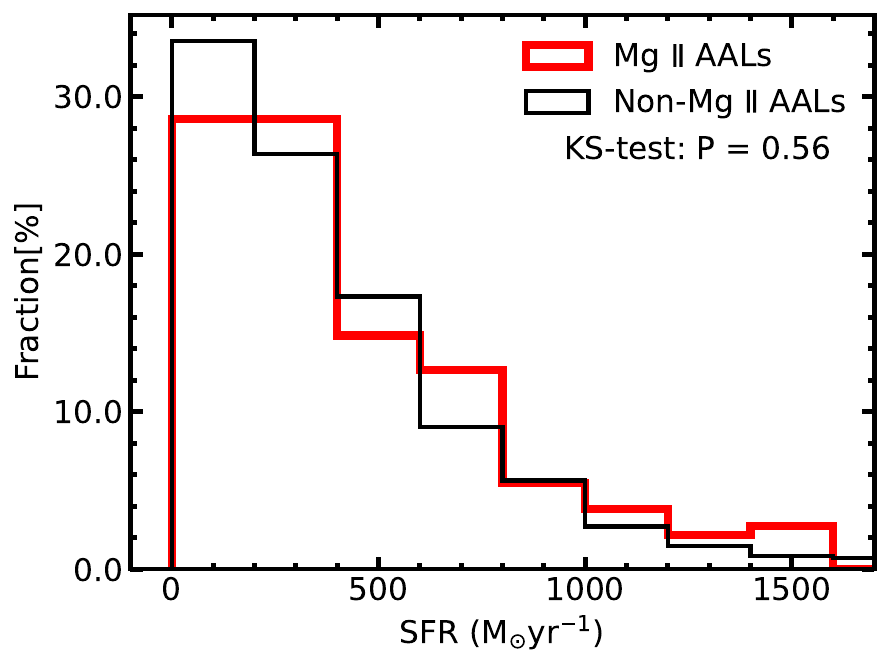}\\
	\caption{The SFR distributions for quasars with (red line) and without (black line) \mgii\ AALs. Here the IR emissions from AGN dust are based on the  template from \cite{2017MNRAS.471...59L}.}
\label{fig:SFR}
\end{figure}

We employ a model with a combination of two components (AGN torus+ star-forming galaxy) to fit the combined  photometrical data (WISE + SPIRE) of each FIR detected quasar in our sample, and estimate their SFRs based on their integrated luminosities at FIR. Figure \textcolor{blue}{\ref{fig:SFR}} shows the distributions of the SFRs for the FIR detected quasars with (red line) and without (black line) \mgii\ AALs, with the medians of $335_{\rm-215}^{\rm+505}$ and $320_{\rm-219}^{\rm+465}$ $\rm M_\odot$~yr$^{\rm-1}$ for the quasars with and without \mgii\ AALs, respectively. A two-sided KS-test on the two distributions finds no statistically significant difference between them with a null probability of being drawn from same parent distribution of $p =0.56$.  Further stacking analysis of the 80 FIR undetected quasars in our sample also finds no statistically significant difference in the SFR distribution for both the quasars with \mgii\ AALs and without \mgii\ AALs (see Section \ref{stacking}).  These suggest that the quasars with \mgii\ AALs have indistinguishable FIR-based SFR distribution from those without \mgii\ AALs.

\begin{figure}[ht!]
	\includegraphics[width=0.49\textwidth]{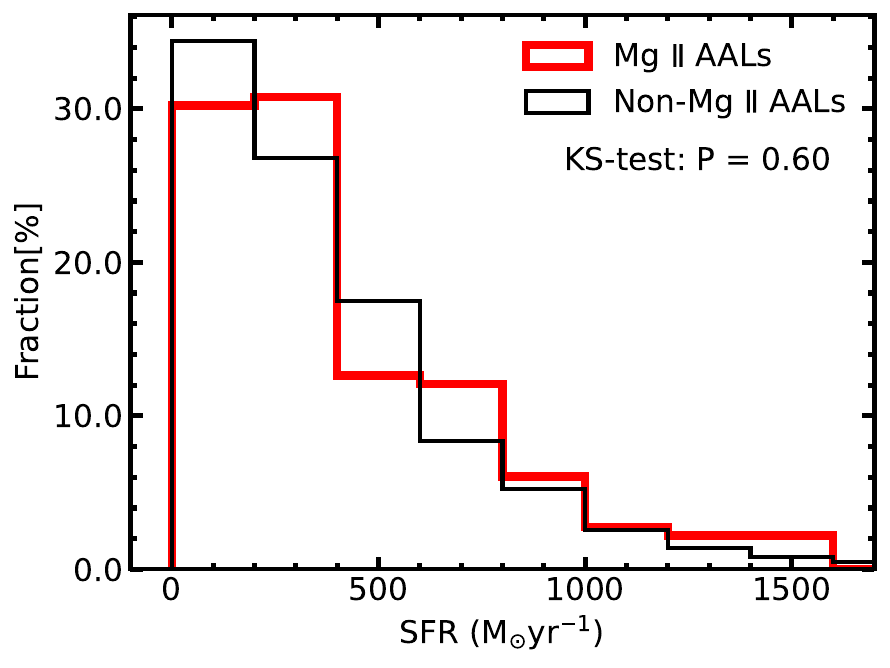}\\
	\caption{The same as Figure {\ref{fig:SFR}} but based on the  template from \cite{2022MNRAS.514.4450S}.}
\label{fig:SFR2022}
\end{figure}

The AGN contributions to the FIR emission are different for different AGN templates, which may lead to a discrepancy in estimates of the SFRs of AGN host galaxies. Different from the AGN template from \cite{2017MNRAS.471...59L}, the AGN template from \cite{2016MNRAS.459..257S,2022MNRAS.514.4450S} has non-negligible contributions to the FIR emission, and powerful AGN tends to have a greater contribution. To assess the effect of AGN template on the SFR distributions of the two types of quasars, we repeat the work as doned in \ref{SED fitting and FIR SFR} by replacing with the template of \cite{2016MNRAS.459..257S,2022MNRAS.514.4450S}. Then the results are shown in Figure \textcolor{blue}{\ref{fig:SFR2022}}. One could find from the figure that, a large p-value ( $p=0.60$) indicates the two types of quasars almost share the same SFR distributions, with the median SFRs are $318_{-203}^{+487}$ and $305_{-209}^{+452}$ $\rm M_\odot ~yr^{-1}$ for the quasars with and without \mgii\ AALs, respectively, which are systematically slightly smaller than those derived from the AGN template of \cite{2017MNRAS.471...59L}. This shoud be the cause that the AGN template of \cite{2016MNRAS.459..257S,2022MNRAS.514.4450S} contributes systematically more to FIR emissions than that of \cite{2017MNRAS.471...59L}.

\begin{figure}[ht!]
	\includegraphics[width=0.48\textwidth]{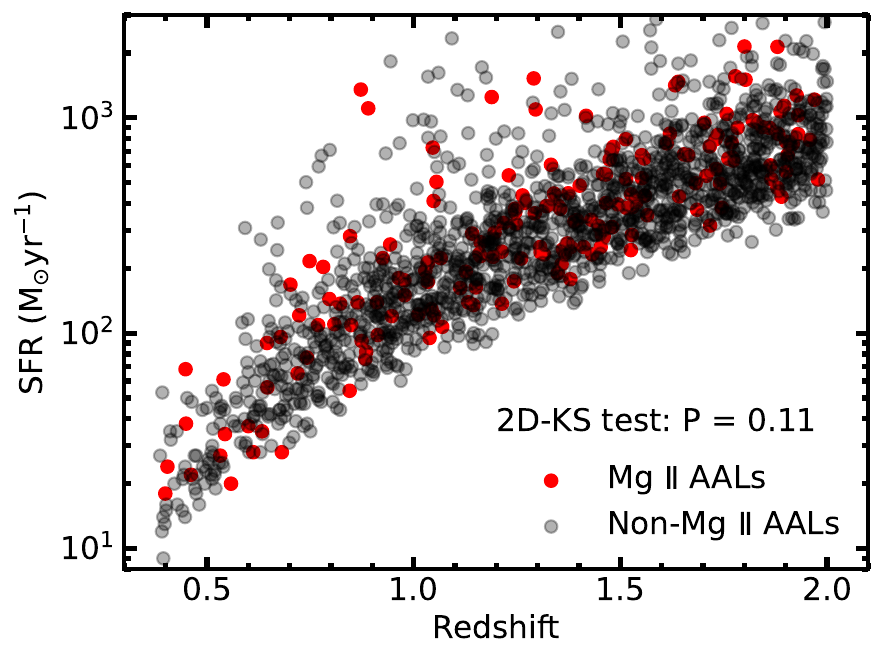}\\
	\caption{The z-SFR distributions for the quasars with (red points) and without (gray points) \mgii\ AAL. The 2D-KS test yields a probability $p= 0.11$.}
\label{fig:z-SFR}
\end{figure}

Interestingly, What accounts for the observed distribution in Figures of \ref{fig:SFR} and \ref{fig:SFR2022} ? Although it is believed that the SFR density is redshift-dependent, with a peak at approximately 3.5 Gyr (z$\sim$2) after the Big Bang \citep{2014ARA&A..52..415M}, we believe that the quasar redshifts in our sample play a less role in producing similar SFR distributions based on the following evidences: firstly, as shown in Figure \ref{fig:z}, the null hypothesis probability ($p=0.25$) returned by a 1-D KS-test indicates that one can not reject the two types of quasars come from the same parent distribution. Next, shown in Figure \ref{fig:z-SFR} are the relationships between the SFRs of the two types of quasars derived from the AGN template of \cite{2017MNRAS.471...59L} and their corresponding redshifts. Again, the null hypothesis probability ($p=0.11$) returned by a 2-D KS-test \citep[][]{1983MNRAS.202..615P} on the SFR-redshift distributions indicates no statistically significant difference between them.

The SFR is also related to its stellar mass within the host galaxy. Although the stellar masses of our sample's quasars  host galaxies are not available from the current literature, we could use the virial black hole masses provided by \cite{2022ApJS..263...42W} and \cite{2023ApJS..264...52H} as a proxy of the stellar masses due to a tight correlation between the two quantities \citep[][]{2013ARA&A..51..511K}. Finally we identify that the median black hole masses of the quasars with and without \mgii\ AALs are $10^{8.79_{-0.48}^{+0.65}}~\rm{M_{\odot}}$ and $10^{8.70_{-0.52}^{+0.50}}~\rm{M_{\odot}}$, respectively. This also suggests that the stellar masses within host galaxies in our sample play a less role in produceing the two indistinguishable FIR-based SFR distributions as shown in Figurs of \ref{fig:SFR} and \ref{fig:SFR2022}.

\begin{figure}[ht!]
	\includegraphics[width=0.45\textwidth]{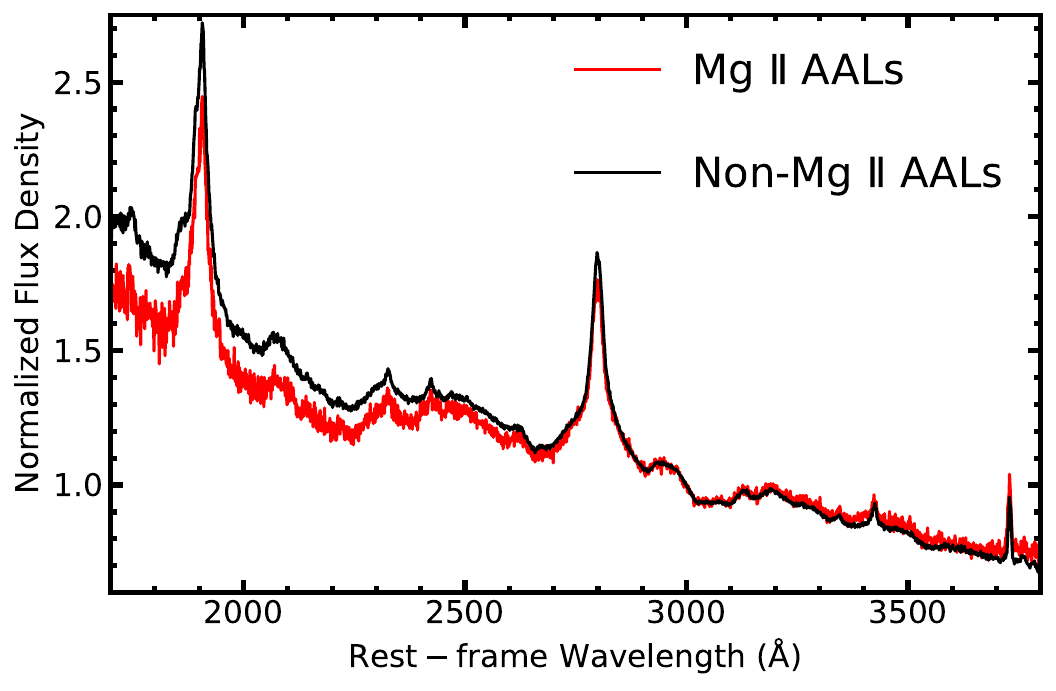}\\
	\caption{The median composite spectra of the Non-\mgii\ AAL (black line) and \mgii\ AAL (red line) quasars.}
\label{fig:composite}
\end{figure}

In the evolution scheme, the quasars with \mgii\ AALs are expected to host redder spectra and higher SFRs relative to those withou \mgii\ AALs \citep[e.g.,][]{2012ApJ...748..131S,2022NatAs...6..339C,2024ApJ...963....3P}. In the orientation-dependent scheme, the quasars with \mgii\ AALs would also exhibit redder spectra relative to those without \mgii\ AALs, but both types of quasars are expected to have similar SFRs. As shown in Figure \textcolor{blue}{\ref{fig:composite}}, our analysis clearly shows that the quasars with \mgii\ AALs obviously exhibit redder median composite spectra compared to those without \mgii\ AALs. However, quasars with \mgii\ AALs are statistically indistinguishable in terms of the SFR distribution from those without \mgii\ AALs (see Figure \ref{fig:SFR} or \ref{fig:SFR2022}). The observations are not supported by the evolutionary model, which puzzle us. Similar observed signatures have also been found between the HiBAL and Non-BAL quasars \citep{2012MNRAS.427.1209C}, in which a simple orientation scheme was invoked to interpret. At the same time, some  authors \citep{2001ApJ...547..635R,2003ApJ...599..116V,2019MNRAS.488.5916S} have also suggested that some AALs are the result of an outflow from the accretion disc, perhaps in a direction at some angle to our line of sight.   Being reminded of this, we speculate that the orientation effect may be a possible channel for reconciling the observations seen in Figure \textcolor{blue}{\ref{fig:composite}}. We will further investigate the properties of the AALs when the angles of quasar's sightlines can be limited with more available data, so that ones can comprehend more deeply the influences of the evolution and orientation effects on the formation of the quasar's AALs.

We also notice that some of the SPIRE fluxes may be contaminated with non-thermal emission, especially for radio-detected quasars. There are only 120 radio-detected quasars in our sample. Thus we exclude these sources from our sample  and still find no statistically significant difference in the SFR distribution for both the quasars with \mgii\ AALs and without \mgii\ AALs, indicatiing that the contamination, if any, do not biase our conclusions.  But a caution is warranted regarding our findings in view of the following facts: The full width half maximum (FWHM) of SPIRE point spread functions (PSF) (beam size) are larger than 18'', reaching 36'' at 500 $\rm\upmu m$ (see SPIRE Handbook\footnote{\url{http://herschel.esac.esa.int/Docs/SPIRE/spire_handbook.pdf}}). These low resolutions lead to a chance that sources physically associated with the quasars, or just the chance line of sight sources, which would contaminate the flux of quasar of interest to us, and now we can not eliminate this contamination with available data. This may bias our findings.

\section*{Acknowledgements}
This work is supported by This work is supported by the Guangxi Natural Science Foundation (2024GXNSF262DA010069), the National Natural Science Foundation of China (grant No. 12133003, 12473011), and the Scientific Research Project of Guangxi University for Nationalities (2018KJQD01). We deeply thank the anonymous referees for her/his helpful and careful comments.
\bibliography{cG}{}
\bibliographystyle{aasjournal}



\end{document}